# Quantum nucleation of effective Minkowski spacetime in hyperbolic metamaterials based on high $T_c$ superconductors

Igor I. Smolyaninov

*Department of Electrical and Computer Engineering, University of Maryland, College Park, MD 20742, USA*

**We demonstrate that high $T_c$ superconductors exhibit hyperbolic metamaterial behavior in the far infrared and THz frequency ranges. In the THz range the hyperbolic behavior occurs only in the normal state, while no propagating modes exist in the superconducting state. Wave equation, which describes propagation of extraordinary light inside a hyperbolic metamaterial exhibits 2+1 dimensional Lorentz symmetry. The role of time in the corresponding effective 3D Minkowski spacetime is played by the spatial coordinate aligned perpendicular to the copper oxide layers. Such superconductor-based hyperbolic metamaterials exhibit a quantum phase transition at T=0, in which the effective Minkowski spacetime arise in the mixed state of the superconductor at some critical value of external magnetic field. Nucleation of Minkowski spacetime occurs via formation of quantized Abrikosov vortices, so that these vortices play the role of effective Minkowski spacetime quanta.**

Propagation of monochromatic extraordinary light inside a hyperbolic metamaterial is described by wave equation, which exhibits 2+1 dimensional Lorentz symmetry [1]. The role of time in the corresponding effective 3D Minkowski spacetime is played by the spatial coordinate, which is oriented along the optical axis of the metamaterial [2]. Since the best description of a physical system must take into account its symmetries, the most appropriate language to describe physics of hyperbolic metamaterials appears to be the language of "effective Minkowski spacetime" [2]. This language establishes connection between hyperbolic metamaterials and other condensed matter analogue models of gravity (see recent extensive review [3] and the references therein, such as [4]). In some sense, hyperbolic metamaterials offer novel and interesting experimental window into physics of Minkowski spacetimes [5-8], which was previously accessible only via theoretical simulations. However, because of the different nature of metamaterial and cosmological systems, there are no direct implications of metamaterial research on gravitational physics.

Despite this obvious limitation, physics of hyperbolic metamaterials appears to be quite novel and interesting. For example, breaking the mirror and temporal (PT) symmetries of the hyperbolic metamaterial may produce one-way propagation of light along the "timelike" spatial coordinate, so that the resulting 2+1 dimensional effective Minkowski spacetime appears to be "causal" [5], while an interface between two different effective Minkowski spacetimes appear to have non-trivial electromagnetic properties [6,7]. These results are even more interesting given the fact that physical vacuum itself may behave as a hyperbolic metamaterial when subjected to very strong magnetic field [9,10]. Here we consider a novel hyperbolic metamaterial system based on high $T_c$ superconductors exhibiting a quantum phase transition at T=0, in which the effective Minkowski spacetime arise at a certain critical value of external magnetic field in the mixed state of the superconductor. The emergent Minkowski spacetime is quantized, since its nucleation occurs via formation of quantized Abrikosov vortices.

As a first step, let us demonstrate that the wave equation describing propagation of monochromatic extraordinary light inside a hyperbolic metamaterial does indeed exhibit 2+1 dimensional Lorentz symmetry. A detailed derivation of this result can be found in [1,2]. We assume that the metamaterial in question is uniaxial and non-magnetic ($\mu$=1), so that electromagnetic field inside the metamaterial may be separated into ordinary and extraordinary waves. Vector $\vec{E}$ of the extraordinary light wave is parallel to the plane defined by the *k*–vector of the wave and the optical axis of the metamaterial. Since hyperbolic metamaterials typically exhibit strong temporal dispersion, we will work in the frequency domain and assume that in some frequency band around $\omega=\omega_0$ the metamaterial may be described by anisotropic dielectric tensor having the diagonal components $\varepsilon_{xx}=\varepsilon_{yy}=\varepsilon_1>0$ and $\varepsilon_{zz}=\varepsilon_2<0$. In the linear optics approximation all the non-diagonal components are assumed to be zero. Propagation of extraordinary light in such a metamaterial may be described by a coordinate-dependent wave function $\psi_\omega=E_z$ obeying the following wave equation [1,2]:

$$-\frac{\omega^2}{c^2}\psi_\omega=\frac{\partial^2\psi_\omega}{\varepsilon_1\partial z^2}+\frac{1}{\varepsilon_2}\left(\frac{\partial^2\psi_\omega}{\partial x^2}+\frac{\partial^2\psi_\omega}{\partial y^2}\right) \qquad (1)$$

(note that the ordinary portion of the electromagnetic field does not contribute to $\psi_\omega$). This wave equation coincides with the Klein-Gordon equation for a massive scalar field $\psi_\omega$ in 3D Minkowski spacetime:

$$-\frac{\partial^2\psi_\omega}{\varepsilon_1\partial z^2}+\frac{1}{(-\varepsilon_2)}\left(\frac{\partial^2\psi_\omega}{\partial x^2}+\frac{\partial^2\psi_\omega}{\partial y^2}\right)=\frac{\omega_0^{\;2}}{c^2}\psi_\omega=\frac{m^{*2}c^2}{\hbar^2}\psi_\omega \qquad (2)$$

in which spatial coordinate $z=\tau$ behaves as a "timelike" variable. For example, it is easy to check that eq.(2) remains invariant under the effective Lorentz coordinate transformation

$$z' = \frac{1}{\sqrt{1-\frac{\varepsilon_1}{(-\varepsilon_2)}\beta}}(z-\beta x) \tag{3}$$

$$x' = \frac{1}{\sqrt{1-\frac{\varepsilon_1}{(-\varepsilon_2)}\beta}}\left(x-\beta\frac{\varepsilon_1}{(-\varepsilon_2)}z\right),$$

where $\beta$ is the effective Lorentz boost. Moreover, similar to our own Minkowski spacetime, together with translations and rotations in the *xy* plane these effective Lorentz transformations form the Poincare group. Thus, eq.(2) describes world lines of massive particles which propagate in a flat 2+1 dimensional Minkowski spacetime [1,2]. Note that the components of metamaterial dielectric tensor define the effective metric $g_{ik}$ of this spacetime: $g_{00}=-\varepsilon_1$ and $g_{11}=g_{22}=-\varepsilon_2$. The metric signature has Minkowski character if $\varepsilon_1$ and $\varepsilon_2$ have opposite signs, and Euclidean character if the signs of $\varepsilon_1$ and $\varepsilon_2$ are the same.

As a second step, let us demonstrate that copper oxide-based high $T_c$ superconductors exhibit hyperbolic metamaterial behavior in the far infrared and THz frequency ranges. Hyperbolic metamaterials are typically composed of multilayer metal-dielectric or metal wire array structures. However, a few natural materials, such as sapphire and bismuth [11] also exhibit hyperbolic behavior in a limited frequency range. Let us demonstrate that typical high Tc superconductors also belong to this class of “natural” hyperbolic materials. Comparison of the crystallographic unit cell of a BSCCO high $T_c$ superconductor and geometry of a layered hyperbolic metamaterial (shown in Fig.1) strongly indicates potential hyperbolic behavior of the high $T_c$ superconductors in an extended spectral range. Indeed, DC conductivity of typical high $T_c$ compounds is known to come from charge carriers (electrons or holes), which mainly reside on the copper oxide planes (Fig.1(a)), while conductivity perpendicular to the copper oxide planes is strongly suppressed [12]. In BSCCO the anisotropy of DC conductivity may reach $10^4$ for the ratio of in plane to out of plane conductivity in high quality single crystal samples. Polarization-dependent AC reflectance spectra measured in the THz and far-infrared frequency ranges [12] also indicate extreme anisotropy. In the normal state of high $T_c$ superconductors the in-plane AC conductivity exhibits Drude-like behavior with a plasma edge close to 10000cm$^{-1}$, while AC conductivity perpendicular to the copper oxide planes is nearly insulating. Extreme anisotropy is also observed in the superconducting state. The typical values of measured in-plane and out of plane condensate plasma frequencies in high $T_c$ superconductors are $\omega_{s,ab}$=4000-10000cm$^{-1}$, and $\omega_{s,c}$=1-1000cm$^{-1}$, respectively [12]. Once again, the measured anisotropy is the strongest in the BSCCO superconductors. Thus, experimental measurements strongly support the qualitative picture of BSCCO structure as a layered hyperbolic metamaterial (Fig.1(b)) in which the copper oxide layers may be represented as metallic layers, while the SrO and BiO layers may be represented as the layers of dielectric. This qualitative picture is supported by the following quantitative analysis.

Let us calculate the diagonal components of the permittivity tensor of a typical high $T_c$ superconductor in the THz and far-infrared ranges using Maxwell-Garnett approximation. These components can be calculated similar to [13] as follows:

$$\varepsilon_1 = \alpha\varepsilon_m + (1-\alpha)\varepsilon_d \quad , \qquad \varepsilon_2 = \frac{\varepsilon_m \varepsilon_d}{(1-\alpha)\varepsilon_m + \alpha\varepsilon_d} \tag{3}$$

where $\alpha$ is the volume fraction of superconducting phase, and $\varepsilon_m<0$ and $\varepsilon_d>0$ are the dielectric permittivities of the superconductor and dielectric, respectively. Note that calculations of permittivity tensor components in ref.[13] has been performed for anisotropic nanoplasmonic composites in the limit of nanometer scale individual layer thicknesses. Therefore, theoretical description developed in ref.[13] is also applicable in our case. However, as has been shown by Larkin and Stockman [14], permittivity of silver and gold may deviate from the predictions of Drude model when particle size (or layer size) becomes smaller than 5 nm. While these corrections may need to be calculated from the first principles, it is reasonable to assume that they will affect the motion of electrons perpendicular to layers stronger than they affect the motion of electrons along the layers.

Our calculations will be performed for $T<T_c$ and centred around the spectral range $\omega_{s,c}<\omega<\omega_{s,ab}$, so that AC conductivity perpendicular to the copper oxide layers may indeed be considered dielectric. For the in-plane permittivity we assume Drude-like behavior supported by AC measurements (see Fig.5 from [12]):

$$\varepsilon_m = \varepsilon_{m\infty} - \frac{\omega_{s,ab}^2}{\omega^2} \quad , \tag{4}$$

while the out of plane permittivity will be approximates as

$$\varepsilon_d = \varepsilon_{d\infty} - \frac{\omega_{s,c}^2}{\omega^2} \quad , \tag{5}$$

where $\varepsilon_{m\infty} \sim 4$ is the dielectric permittivity of copper oxide layers above the plasma edge, and $\varepsilon_{d\infty}$ is the dielectric permittivity of the undoped insulating parent copper oxide compound. The parent perovskite compounds typically exhibit rather large values of $\varepsilon_{d\infty}$, which may be estimated from their out of plane reflectivity. Based on Fig.5 from [12] $\varepsilon_{d\infty} \sim 25$ may be assumed. The calculated diagonal components of the permittivity tensor for a high $T_c$ superconductor having $\omega_{s,ab}$=10000cm$^{-1}$ and $\omega_{s,c}$=1000cm$^{-1}$ are presented in Fig.2. Based on the crystallographic unit cell of BSCCO $\alpha$=0.36 has been assumed in these calculations. The hyperbolic behavior appears in the 200cm$^{-1}$<$\omega$<1200cm$^{-1}$ and 2400cm$^{-1}$<$\omega$<4800cm$^{-1}$ spectral ranges. The appearance of hyperbolic bands is quite generic, independent of a particular choice of $\omega_{s,ab}$ and $\omega_{s,c}$ , as long as strong anisotropy of $\omega_s$ is maintained. Within these spectral bands $\varepsilon_1$ and $\varepsilon_2$ have opposite signs, so that the effective spacetime described by eq.(2) has Minkowski character. Appearance of such an effective Minkowski spacetime has clear experimental signatures. If a dipole source oscillating at a frequency $\omega_0$ located within the hyperbolic band is placed inside the high $T_c$ superconductor, its radiation pattern looks like a light cone in Minkowski spacetime [5] (see inset in Fig.2).

As a side note, we should point out that the demonstrated hyperbolic character of photon dispersion in high $T_c$ superconductors may have important consequences for the electron pairing interaction. Importance of plasmon modes for electron pairing in layered high $T_c$ superconductors has been emphasized by many authors in the past (see for example [15]). However, the fact that hyperbolic character of photon dispersion leads to appearance of broadband divergences in the photonic (plasmonic) density of states has been realized only very recently [1, 16]. This broadband divergence (which is cut off at very large scale $k$~3 nm$^{-1}$ corresponding to distance between copper oxide layers) should lead to considerable enhancement of plasmon-mediated pairing of electrons proposed in [15]. Another potentially important consequence of hyperbolic behavior of high Tc superconductors is that these materials may be used in hyperlens-based superresolution imaging [17,18] in the THz and far-infrared frequency ranges. Such novel hyperlenses will have very interesting switchable properties as a function of temperature, current and external magnetic field.

Let us now investigate if a metric signature transition [1] may occur upon transition from superconducting to normal state of a high $T_c$ superconductor. Based on experimental data reported in Fig.5 from ref.[12] it appears that the far-infrared hyperbolic band at 2400cm$^{-1}$<$\omega$<4800cm$^{-1}$ is not affected by the superconducting transition. This band is primarily defined by the plasma edge of the in-plane AC conductivity located close to 10000cm$^{-1}$. This plasma edge is mostly unaffected by the superconducting transition. On the other hand, as demonstrated in Fig.3, the lower frequency hyperbolic band at $\omega$<1200cm$^{-1}$ is strongly affected by the transition from superconducting to normal state. Calculations performed using the same parameters as in Fig.2 indicate that a metric signature transition is observed for $\omega$<180cm$^{-1}$. This transition is caused by the disappearance of the $\omega_{s,c}$ term in eq.(5) in the normal state, which is observed in the experiment [12]. Thus, in the superconducting state photons cannot propagate below $\omega$<180cm$^{-1}$ since $\varepsilon$ is negative in all directions. Upon transition into normal state, an effective 3D Minkowski spacetime appear for $\omega$<180cm$^{-1}$ extraordinary photons from the superconducting "nothingness". Such a transition may occur at $T$=0 as a function of external magnetic field. Therefore it belongs to a class of quantum phase transitions. A somewhat similar quantum metric signature transition may occur at $T$=0 in physical vacuum subjected to a much stronger magnetic field [9,10]. Note that the measured room temperature in-plane reflectivity of the typical high $T_c$ superconductors below $\omega$<180cm$^{-1}$ exceeds 94% (see Fig.5 from ref.[12]). Therefore, Im($\varepsilon_1$)/Re($\varepsilon_1$), which is dominated by Im($\varepsilon_m$)/Re($\varepsilon_m$) does not exceed 0.06 even at room temperature and further decreases on cooling, making high $T_c$ superconductors an excellent choice for studying emergence of effective Minkowski spacetimes in hyperbolic metamaterials.

Let us consider the detailed mechanism of this transition from the point of view of extraordinary photons in the $\omega$<180cm$^{-1}$ spectral range. At $H<H_{c1}$ magnetic field does not penetrate into the superconductor, and no propagating electromagnetic modes exist in the superconducting state. We could characterize this state as "electromagnetic nothingness". On the other hand, as demonstrated by Fig.3, at $H>H_{c2}$ the high $T_c$ superconductor will be in the normal "hyperbolic metamaterial" state. As described earlier, extraordinary photons in the $\omega$<180cm$^{-1}$ spectral range perceive this state as an effective 3D Minkowski spacetime. “Nucleation” of this effective Minkowski spacetime from superconductive “nothingness” occurs via the mixed state, in which the external magnetic field gradually penetrates into the superconductor via formation of quantized Abrikosov vortices [19]. These vortices have normal cores with dimensions approximately equal to the superconducting coherence length $\xi$. Therefore, they may be considered as elementary “Minkowski spacetime quanta”. Even though $\xi$ is much smaller than the free space wavelength $\lambda$ in the $\omega$<180cm$^{-1}$ spectral range, the macroscopic electrodynamics treatment of vortices as small pieces of hyperbolic metamaterial makes sense due to divergent nature of photon dispersion law in a hyperbolic metamaterial:

$$\frac{\omega^2}{c^2} = \frac{k_z^2}{\varepsilon_1} + \frac{k_x^2 + k_y^2}{\varepsilon_2} \tag{6}$$

This divergence (due to opposite signs of $\varepsilon_1$ and $\varepsilon_2$) is cut off only when the photon $k$ vector components are comparable to the inverse crystallographic cell size [1,16]. When the external magnetic field is gradually increased above $H_{c1,}$ the individual Abrikosov vortices form a crystal lattice. This lattice of “Minkowski spacetime quanta” may be considered as a homogeneous medium once the Abrikosov lattice periodicity becomes much smaller than $\lambda$. The volume fraction of normal phase in the mixed state may be obtained as [19]

$$\beta = \frac{\xi^2 H}{\Phi_0} \approx \frac{H}{H_{c2}} \tag{7}$$

where $\Phi_0$ is the magnetic flux quantum. Using eqs.(3-5,7) we may calculate the averaged diagonal components of the permittivity tensor in the mixed state as a function of external magnetic field. These calculations are performed by taking into account the increase in volume fraction of normal phase given by eq.(7), which reduces $\omega_{s,ab}$ in

eq.(4). Results of these calculations performed at $\omega$=150cm$^{-1}$ are presented in Fig.4. These calculations were performed using the same parameters as in Fig.2. Macroscopic effective Minkowski spacetime emerges from the Abrikosov vortex lattice at the critical field $H_M$=0.47$H_{c2}$. Emergence of macroscopic Minkowski domains within the Abrikosov lattice state also has clear experimental signatures. As demonstrated in the inset in Fig.4, electromagnetic field at the Minkowski domain boundaries diverges in the zero loss limit. As has been demonstrated in refs.[6,20], from the electromagnetic standpoint these boundaries are somewhat similar to the black hole and cosmological event horizons. Metamaterial losses, which are unavoidable in hyperbolic metamaterials tame these divergences. However considerable field enhancement is still observed at Im($\varepsilon_1$)/Re($\varepsilon_1$)~0.06 used in our simulations. Thus, the described system may indeed be used as an experimental model of hypothesized quantum nucleation of Minkowski spacetime from “nothing”. An interesting additional feature of this model is that transition from quantum to continuous Minkowski spacetime may be traced as a function of external magnetic field.

In conclusion, we have demonstrated that high Tc superconductors exhibit hyperbolic metamaterial behavior in the far infrared and THz frequency ranges. Such superconductor-based hyperbolic metamaterials exhibit a quantum phase transition at T=0, in which the effective Minkowski spacetime arise in the mixed state of the superconductor at some critical value of external magnetic field. Nucleation of Minkowski spacetime occurs via formation of quantized Abrikosov vortices, so that these vortices play the role of Minkowski spacetime quanta.

I acknowledge helpful conversations with Vera Smolyaninova.

**Figure Captions**

**Figure 1.** Comparison of the crystallographic unit cell of a BSCCO high $T_c$ superconductor (a) and geometry of a layered hyperbolic metamaterial (b).

**Figure 2**. Diagonal components of the permittivity tensor of a high Tc superconductor calculated as a function of frequency assuming $\omega_{s,ab}$=10000cm$^{-1}$, $\omega_{s,c}$=1000cm$^{-1}$, $\varepsilon_{m\infty} = 4$ and $\varepsilon_{d\infty} = 25$. The hyperbolic bands appear at 200cm$^{-1}$<$\omega$<1200cm$^{-1}$ and 2400cm$^{-1}$<ω<4800cm$^{-1}$. The inset shows radiation pattern of a dipole source placed inside a hyperbolic metamaterial calculated using COMSOL Multiphysics 4.2 solver. Radiation pattern looks like a light cone in a 2+1 dimensional Minkowski spacetime in which spatial z coordinate plays the role of a "timelike" coordinate.

**Figure 3**. Comparison of diagonal components of the permittivity tensor of a high Tc superconductor in superconducting and normal states in the $\omega$<1200cm$^{-1}$ frequency range. Calculations were performed using the same parameters as in Fig.2. The metric signature transition is observed for $\omega$<180cm$^{-1}$.

**Figure 4.** Diagonal components of the permittivity tensor of a high Tc superconductor calculated at T=0 as a function of external magnetic field at $\omega$=150cm$^{-1}$. Calculations were performed using the same parameters as in Fig.2. Macroscopic effective Minkowski spacetime emerge from the Abrikosov vortex latice at $H_M$=0.47$H_{c2}$. The inset shows an example of electromagnetic field distribution in a random configuration of Minkowski domains calculated using COMSOL Multiphysics 4.2 solver. In the limit of zero losses electromagnetic field diverges at the Minkowski domain boundaries (the shown simulation assumes Im($\varepsilon_1$)/Re($\varepsilon_1$)~0.06).

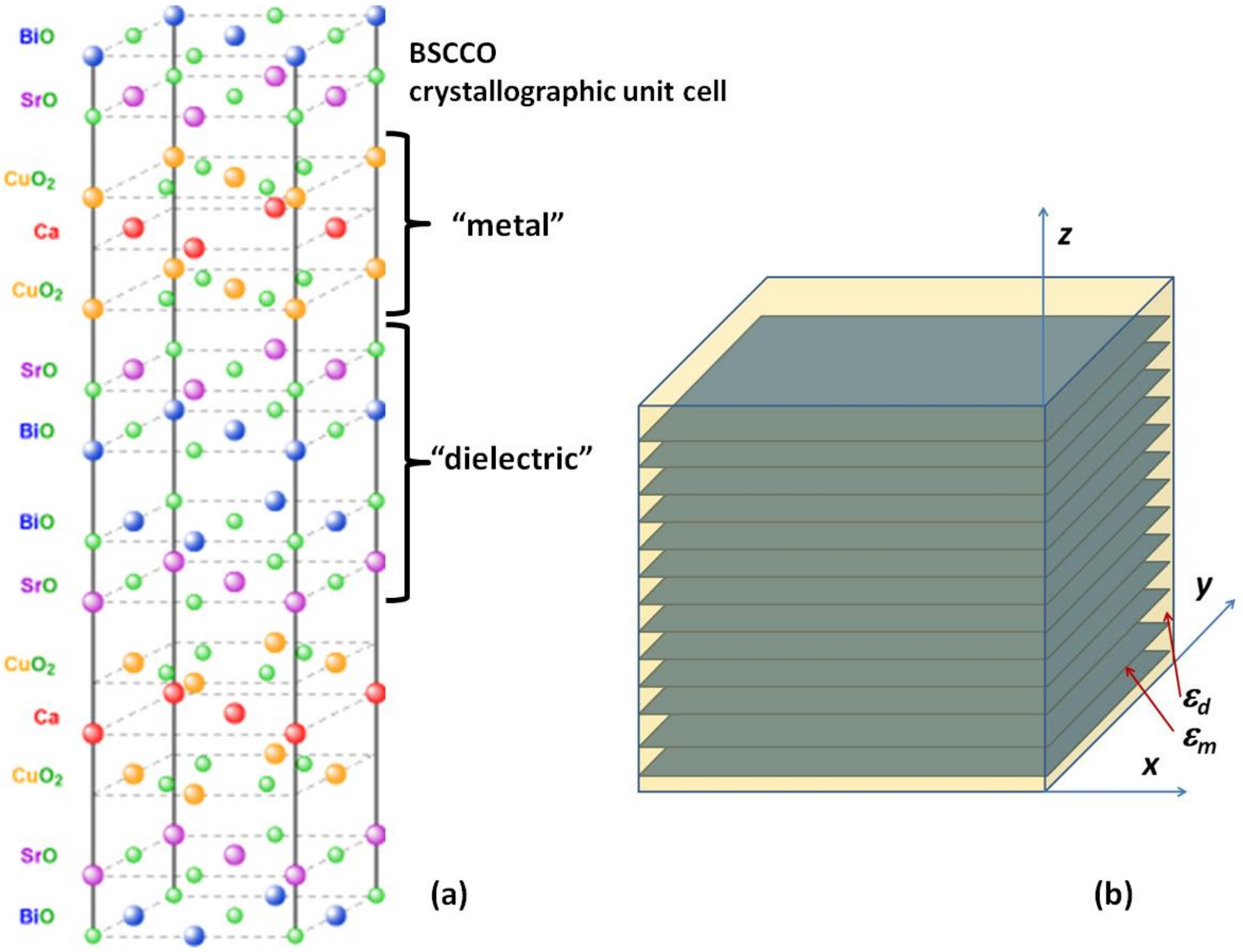

Fig.1

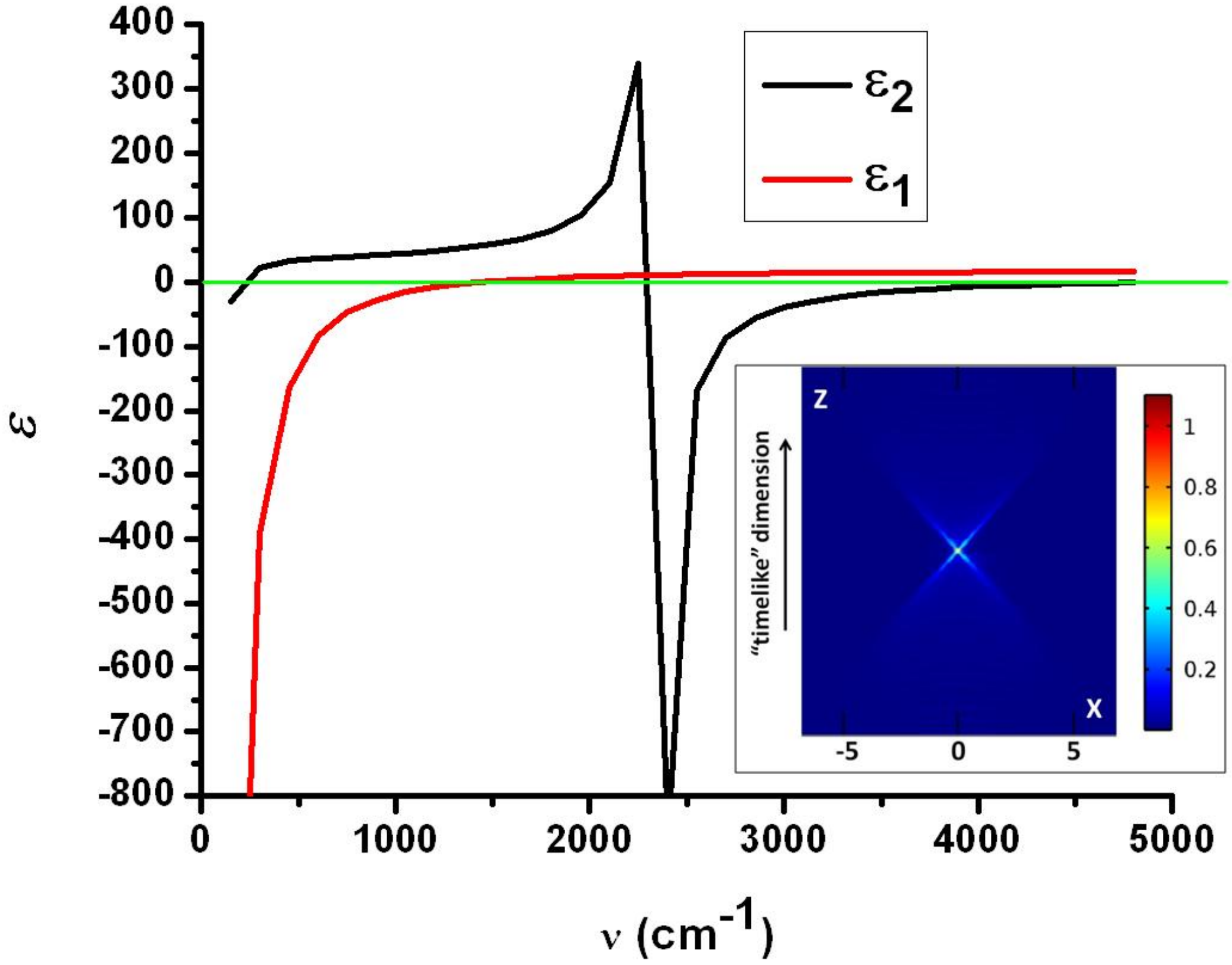

Fig.2

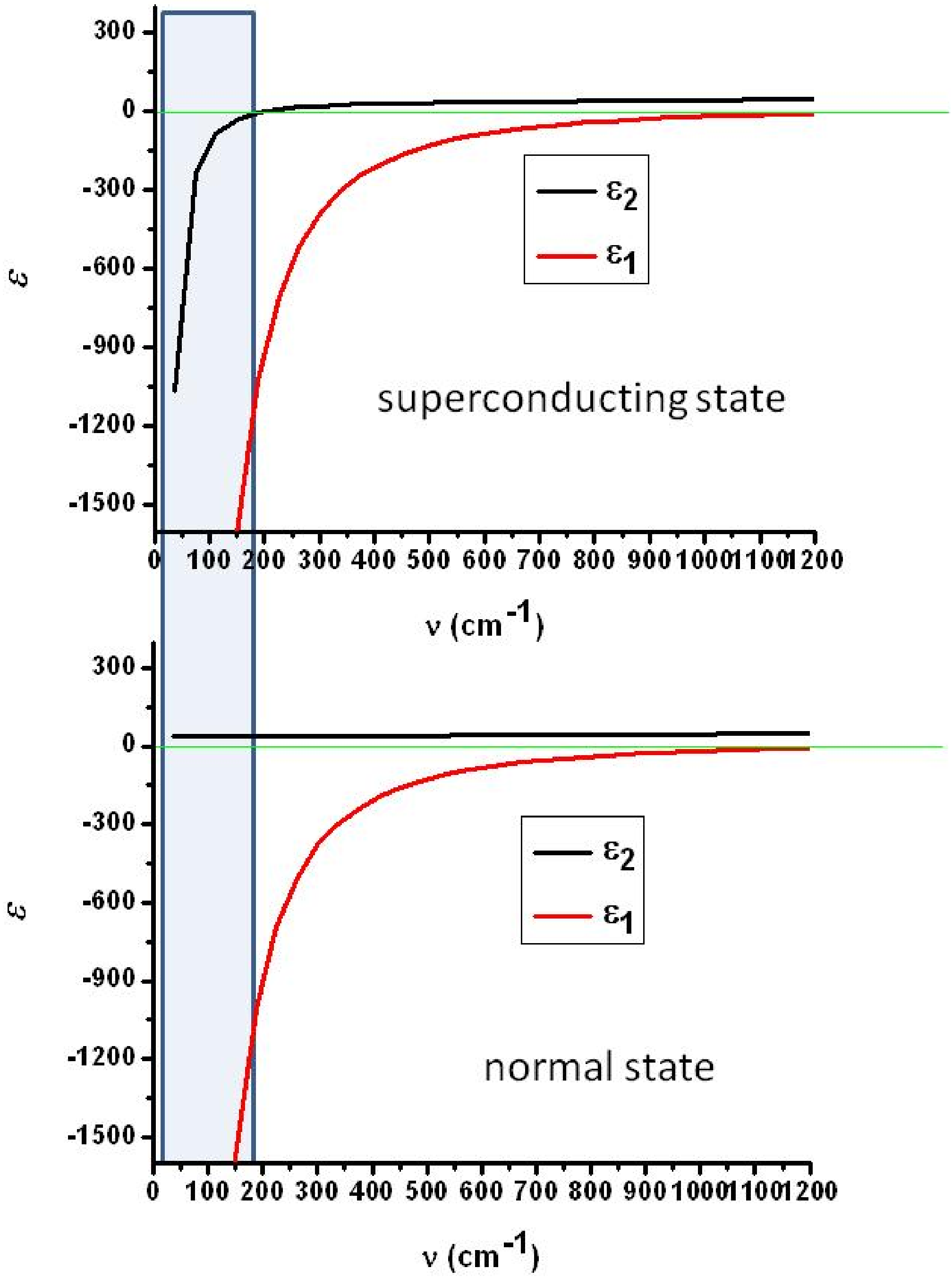
300
0
-300
-600
-900
-1200
-1500
ε
ε2
ε1
superconducting state
0 100 200 300 400 500 600 700 800 900 1000 1100 1200
ν (cm-1)
300
0
-300
-600
-900
-1200
-1500
ε
ε2
ε1
normal state
0 100 200 300 400 500 600 700 800 900 1000 1100 1200
ν (cm-1)

Fig.3

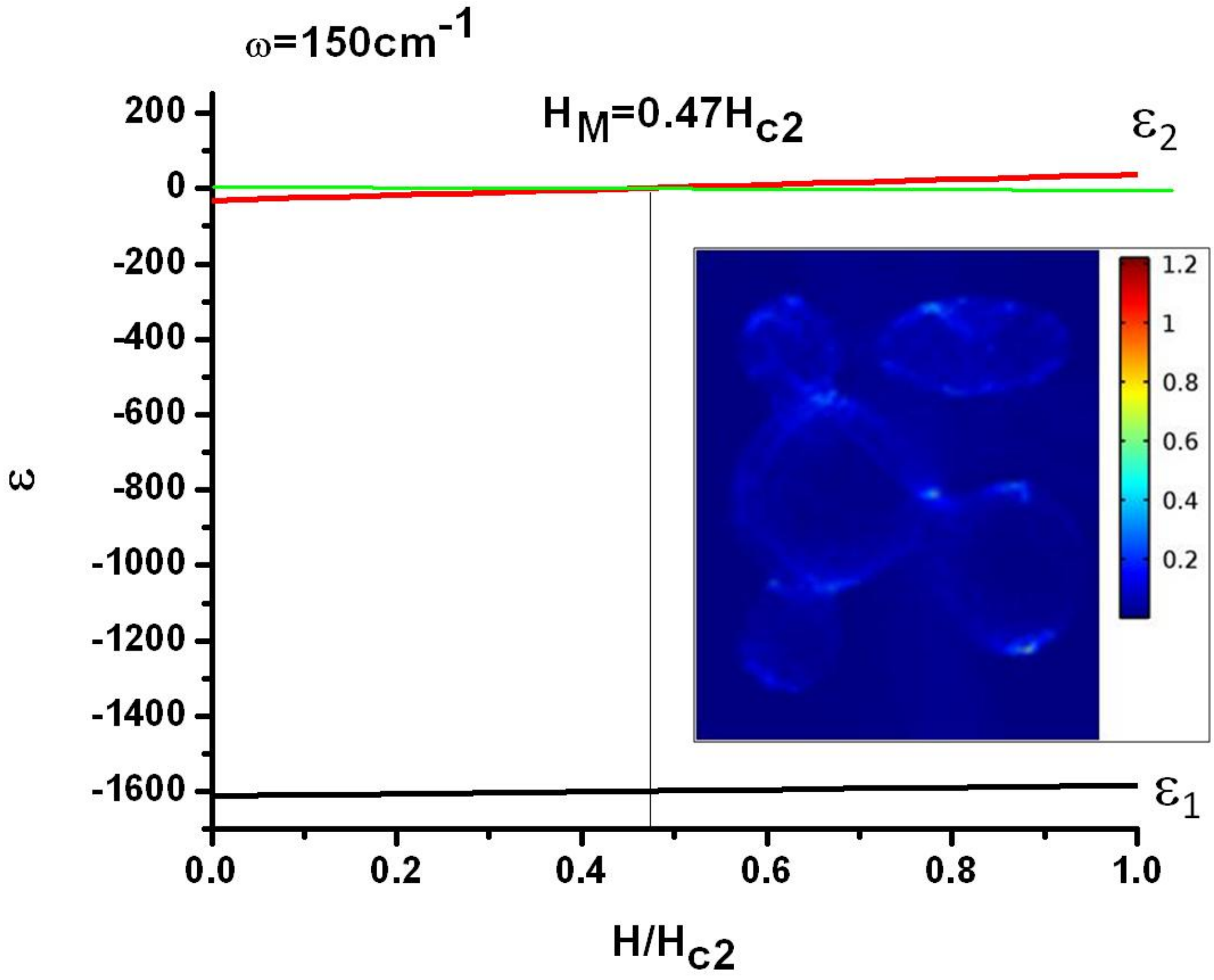

Fig.4